\documentstyle[12pt]{article}

\newcommand{\eq}{\begin{equation}}
\newcommand{\en}{\end{equation}}

\newcommand{\eqa}{\begin{eqnarray}}
\newcommand{\ena}{\end{eqnarray}}
\newcommand{\eqan}{\begin{eqnarray*}}
\newcommand{\enan}{\end{eqnarray*}}

\newcommand{\lbl}{\label}

\newcommand{\JAP}[1]{J. Appl. Phys.\ {\bf #1}\ }

\newcommand{\JMaP}[1]{J. Math. Phys.\ {\bf #1}\ }
\newcommand{\JP}[1]{J. Phys.\ {\bf #1}\ }

\newcommand{\PR}[1]{Phys. Rev\ {\bf #1}\ }

\newcommand{\RNC}[1]{Riv. Nuovo Cimento\ {\bf #1}\ }

\def\sqr#1#2{{\vcenter{\hrule height.#2pt
     \hbox{\vrule width.#2pt height#1pt \kern#1pt
        \vrule width.#2pt}
     \hrule height.#2pt}}}

\def\thinspace{\kern .16667em}

\def\Dir{\nabla\kern-7.8pt\Big{/}}

\def\reali{{\hbox{\s@ l\kern-.5mm R}}}
\def\naturali{{\hbox{\s@ l\kern-.5mm N}}}
\def\interi{{\mathchoice
 {\hbox{Z\kern-1.5mm Z}}
 {\hbox{Z\kern-1.5mm Z}}
 {\hbox{{Z\kern-1.2mm Z}}}
 {\hbox{{Z\kern-1.2mm Z}}}  }}

\def\unity{{\hbox{\s@ 1\kern-.8mm l}}}
\def\uno{{\hbox{ 1\kern-.8mm l}}}
\def\pd#1{{\partial~\over\partial #1}}

\def\part{\partial}

\def\aa{\alpha}

\def\dd{\delta}
\def\DD{\Delta}
\def\ee{\epsilon}
\def\eb{\bar\epsilon}

\def\ff{\phi}

\def\gg{\gamma}
\def\GG{\Gamma}

\def\ss{\sigma}

\def\G{{\cal G}}

\begin{document}
\begin{titlepage}
\begin{flushright}
NBI-HE-93-16\\
March 1993\\
hep-lat/9303015
\end{flushright}
\vspace*{0.5cm}
\begin{center}
{\bf
\begin{Large}
{\bf
THE DIMER PARTITION FUNCTION
\\}
\end{Large}
}
\vspace*{1.5cm}
         {\large Igor Pesando}\footnote{E-mail PESANDO@NBIVAX.NBI.DK,
22105::PESANDO, 31890::I\_PESANDO}
         \\[.5cm]
          The Niels Bohr Institute\\
          University of Copenhagen\\
          Blegdamsvej 17, DK-2100 Copenhagen \O \\
          Denmark
\end{center}
\vspace*{0.7cm}
\begin{abstract}
{We apply the Ginzburg criterion to the dimer problem and we solve the
apparent contradiction of a system with mean field $\alpha={1\over2}$,
the typical value of tricritical systems, and upper critical dimension
$D_{cr}=6$.
We find that the system has upper critical dimension $D_{cr}=6$ , while for
$D\le4$ it should undergo a first order phase transition.
We comment on the latter wrong result examining the approximation we used.
}
\end{abstract}
\vfill
\end{titlepage}

\setcounter{footnote}{0}
\def\ut{{\tilde u}}
\def\zt{{\tilde z}}
\def\dz{{\sqrt{2}z}}

\def\uij{U_{ij}}
\def\ucij{U^\dagger_{ij}}
\def\uji{U_{ji}}
\def\ucji{U^\dagger_{ji}}
\def\dag{\dagger}
\def\V{{\cal V}}
\def\S{{\cal S}}
\def\zpm{z_{\pm}}
\def\zp{z_+}
\def\zm{z_-}
\def\ddt{{\dd T}}
\def\mucr{\mu_{cr}}

In this letter we would like to show how it is possible to recover the
upper critical dimension of the dimer system without using
renormalization group arguments but with the use of the Ginzburg
criterion.
In this way we solve what could appear as a contradiction: the mean
field critical
exponent $\aa={1\over2}$ and the upper critical dimension $D_{cr}=6$;
in fact $\aa={1\over2}$ is the typical value of the mean field
critical exponent of the
tricritical transitions  that have upper critical dimension 3.
Differently from previous works (\cite{Sh,KW}) we do not use renormalization
group arguments.

We consider the following action defined on a lattice $\G$
\footnote{
We use $i,j,\dots$ to indicate sites (vertices) of the lattice.
}
\cite{Sa}
\eq
Z_{dimer}=
\int~\prod_{l\in\G} d\eb_l~d\ee_l~
\exp{\left(-\mu\sum_i\eb_i\ee_i
           -{K\over2}\sum_{i,j}A_{i j}\eb_i\ee_i\eb_j\ee_j
     \right)}
\lbl{z1}
\en
where $A_{i j}$ is the adjacency matrix of the lattice $\G$,
$\eb_i=(\ee_i)^*$ are complex grassman variables defined on the vertex
$i$.
It is easy to show that the partition function (\ref{z1}) is the
generating function for the the dimer problem with negative activity,
more explicitly:
\eq
Z_{dimer}=
\mu^\V\sum_D~N(D)~(-{K\over\mu^2})^D
\en
where $D$ is the number of dimers and $N(D)$ is the number of possible
configurations with $D$ dimers ($\V$ is the number of lattice sites).

To prove this assertion we consider the high temperature expansion
(HTE) in the variable ${K\over\mu^2}\sim{1\over T}$.
Now if a term of the HTE gives a non vanishing contribution to the
partition function $Z$, it must have at most one active link (dimer) per each
vertex since we cannot put two dimers on the lattice sharing a vertex
because of the relation $(\eb_i\ee_i)^2=0$.
\noindent
As an observation we want to notice that setting $\mu=0$
we can get the close packing dimer problem that it is known to be
equivalent to Ising and hence, in this case, the theory has critical
dimension $D_{cr}=4$ (\cite{FiRR})\footnote{
To this purpose we have to introduce the terminal lattice of the expanded
version of the original lattice, then we divide the vertices into two
sets: the first one ${\cal G}_1$ containing the vertices of the
original lattice and the other ${\cal G}_2$ with the new vertices.
Finally we generalize the kinetic term as follows:
$$
A_{ij}\Longrightarrow
{\tilde A}_{i j}=\left\{\begin{array}{ll}
                        J & \mbox{if $i$ and $j$ are neighbours
                                   and both belong to lattice ${\cal G}_1$}\\
                        1 & \mbox{if $i$ and $j$ are neighbours}
                        \end{array}
                 \right.
$$
then there exists a critical value of $J$ at which a second order phase
transition occurs.
}.

Before we can use Ginzburg criterion, we must rewrite the partition
function  (\ref{z1})  in a suitable form for the application of the
saddle point expansion.
To this purpose we introduce the variable $R_i=\sqrt{K}\eb_i\ee_i$ and
we use the identity
\eq
\int~ d\eb~d\ee~\dd(R-\sqrt{K}\eb\ee)=\sqrt{K}{d \dd(R)\over d r}
\lbl{delta}
\en
in order to rewrite (\ref{z1}) as follows
$$
Z_{dimer}=
\left.
(-\sqrt{K})^\V~\left(\prod_{k\in\G} \pd{R_k}\right)
\exp{\left(
-{\mu\over\sqrt{K}}\sum_i R_i-{1\over2}\sum_{i,j}A_{i j}R_iR_j
\right)}
\right|_{R_k=0}
$$
If we perform the substitutions $R_i\rightarrow-R_i$ and
${\mu\over\sqrt{K}}\rightarrow\mu$, we use the fact that the derivatives
act on an analytic function, we can finally rewrite the previous
equation as
\eq
Z_{dimer}=(\sqrt{K})^\V
\oint_{\GG}
\prod_{k\in\G} {dz_k\over 2\pi i}
\exp{\left(\sum_i \mu z_i-2\log(z_i)-{1\over2}\sum_{i,j}A_{i j}z_i
z_j\right)}
\lbl{z}
\en
where $\GG$ is an hypersurface surrounding the origin.

We introduce the notation
\eqa
\S&=&\sum_i \mu z_i-2\log(z_i)-{1\over2}\sum_{i,j}A_{i j}z_iz_j
\nonumber\\
\S_{i}&=&-{2\over z_i}+\mu-\sum_j A_{i j}z_j
\nonumber\\
\S_{i j}&=&{2\over z_i^2}\dd_{i j}-A_{i j}
\nonumber
\ena
Under the hypothesis of a translationally invariant solution,
the saddle point equations $\S_i=0$ yield the solutions
\eq
\zpm={\mu\pm\sqrt{\mu^2-8q}\over 2q}
\en
where for $\mu>0$ $\zm$ is a minimum and $\zp$ is a maximum and
the critical value is attained for $\mu^2_{cr}=8q$ where the two
solutions coalesce.

Let us now compute the critical exponent $\aa$. It turns out that in the mean
field approximation this critical exponent is independent on the
choice of $\zp$ or $\zm$, nevertheless we must choose $\zm$ (the minimum); an
explanation of that is delayed to the computation of the first
loop corrections.
It is easy to obtain
\eq
F^{(0)}={\log{Z^{(0)}}\over\V}=-
\mu\zm-2\log(\zm)-{q\over2}\zm^2
\en
and if we identify $T=\mu^2$, we also get
\eqa
E^{(0)}&=&{1\over2}\mu^3\zm
\nonumber\\
C^{(0)}&=&{3\over4}\mu\zm+{1\over4}\mu^2{d \zm\over d\mu}
\ena
It is now immediate to find the behaviour at the critical point
($\ddt=\mu^2-\mucr^2=\mu^2-8q$):
\eqa
E^{(0)}&=& 2\mucr^2-2\mucr\sqrt{\ddt}+O(\ddt)
\nonumber\\
C^{(0)}&=&-{\mucr\over\sqrt{\ddt}}+O(1)
\lbl{alfa0}
\ena
Form these equations we can read immediately that the mean field value
$\aa={1\over2}$ as it should be.

Let us now turn to the exam of the first loop correction.
In order to decide which of the two stationary points we must choose,
we use the observation that if we cannot find
a proper path $\gg$ for the function $S(z)=\S|_{z_i=z}$, we surely cannot find
a proper hypersurface $\GG$.
Now since  the path $\gg$ has to surround the origin and $\zp$ is a maximum,
we cannot find a proper path $\gg$ for $S(z)$ that crosses its saddle but
only paths that raise and descend the saddle on the same side, it
follows that there is no proper path $\gg$ and hence no proper $\GG$.
It follows that we must choose $\zm$; it is not difficult to check that
the path $\gg(t)=\zm e^{i t}$ with $0\le t<2\pi$ has all the
characteristics necessary for the application of the saddle point
method.

If we compute the first loop correction on an hypercubic lattice, we find
\eq
F^{(1)}={1\over2}log{\pi}
-{1\over2}\int_{-\pi}^{\pi}~{d^D p\over (2\pi)^D}
  ~log\left({2\over\zm^2}-2\sum_\nu cos(p_\nu)\right)
\en
from which we can compute both the correction to the energy and to the
specific heat due to the first loop correction, explicitly we have
\eqa
E^{(1)}&=&-\left({\mu\over\zm}\right)^3
          \,{d\zm\over d\mu} \,{d F^{(1)}\over d(1/\zm^2)}=
\nonumber\\
&=&-\left({\mu\over\zm}\right)^3{1\over q}
    \left(1-{2\mu\over\sqrt{\mu^2-\mucr^2}}\right)
    \int_{-\pi}^{\pi}~{d^D p\over (2\pi)^D}
  ~{1\over {2\over\zm^2}-2\sum_\nu cos(p_\nu)}
\ena
and
\eqa
C^{(1)}&=&
-{3 \mu\over 2\zm^3} {d\zm\over d\mu}
  {d F^{(1)}\over d {1\over\zm^2}}
+{2\mu^2\over\zm^4} \left( {d\zm\over d\mu}\right)^2
  {d F^{(1)}\over d {1\over\zm^2}}
-{\mu^2\over2\zm^3} {d^2\zm\over d\mu^2}
  {d F^{(1)}\over d {1\over\zm^2}}
-{\mu^2\over2\zm^3} {d\zm\over d\mu}
  {d^2 F^{(1)}\over d\left(1\over\zm^2\right)^2}=
\nonumber\\
&=&-{\mu^4\over q\zm^3}
    {1\over\sqrt{\mu^2-\mucr^2}^3}
    \int_{-\pi}^{\pi}~{d^D p\over (2\pi)^D}
  ~{1\over {2\over\zm^2}-2\sum_\nu cos(p_\nu)}
\nonumber\\
&&
-{2\mu^2\over q\zm^3}
    {1\over\sqrt{\mu^2-\mucr^2}}
    \int_{-\pi}^{\pi}~{d^D p\over (2\pi)^D}
  ~{1\over ({2\over\zm^2}-2\sum_\nu cos(p_\nu) )^2}+\dots
\ena
Using the fact that
${1\over\zm^2}-{1\over z_{-cr}^2}\sim{2q\over\mucr^2}\sqrt{\ddt}$
we easily deduce the critical behaviour of these quantities:
\eqa
F^{(1)}&\sim&(\sqrt{\ddt})^{{D\over2}}\log(\ddt)
\nonumber\\
E^{(1)}&\sim&(\sqrt{\ddt})^{{D\over2}-2}
\nonumber\\
C^{(1)}&\sim&(\sqrt{\ddt})^{{D\over2}-4}
\lbl{alfa1}
\ena
{}From the first equation we deduce immediately that we should have a first
order
transition for $D\le4$, while from the comparison between the second one
and the second of (\ref{alfa0}) we get the condition
${D\over2}-4>-1$, i.e. $D>6$.

Now we would comment about the rigour of these results.
The first observation is that we applied the saddle point expansion
without any "large" parameter and hence we should keep in account also the
higher order, as we will show the higher order give milder divergences
on $\aa$ provided $D\ge6$.
To show this we notice that the coefficient of a generic $n$-vertex is
$\sim {1\over\zm^n}$ and that the generic contribution to the $L$ loop
correction to $F^{(L)}$ for $\mu\sim\mucr$ is
\eq
F^{(L)}\sim
\prod_{n\ge3} \left({1\over\zm^n}\right)^{V_n}
\int ~d^D p_1\dots d^D p_L
{1\over \DD({1\over\zm^2})+q_1^2}\dots{1\over \DD({1\over\zm^2})+q_I^2}
\en
where $V_n$ is the number of $n$ vertices and $I$ is the number of
internal lines.
Using $\DD({1\over\zm^2})\sim \sqrt{\ddt}$, we get immediately
\eqa
F^{(L)}&\sim&
 {\sqrt{\ddt}^{{D L\over2}-I}\over\zm^I}
\nonumber\\
E^{(L)}&\sim& \sqrt{\ddt}^{{D L\over2}-I-1}
\nonumber\\
C^{(L)}&\sim& \sqrt{\ddt}^{{D L\over2}-I-2}
\ena
With the help of $2I=\sum_n n V_n$ (the graphs have no external legs)
and of $L=I-\sum_n V_n +1$, we find
$D L-2I=D+{1\over2}\sum_{n\ge3}(D(n-2)-2n)V_n$
and hence for $D\ge6$ it follows that $D L-2I\ge D$.
Finally we get $-\aa^{(L)}\ge{D\over2}-2\ge1$ and similarly we find that
the energy cannot diverge.
All this means that the method is selfconsistent and that for $D<6$
the fluctuation are important and the renormalization group is needed.
As far as the dimension where the first order phase transition takes
place, we cannot be sure that the found value $4$ is the correct one
because from the previous discussion we know that this value is deep inside the
region where the analysis is not more reliable.
The proof of the fact that in $D=4$ {\sl no} first order phase transition
takes place was done long time ago in (\cite{KW}) in connection to the
Yang-Lee edge singularity.
In this work it was performed an analysis of the
scaling dimension of the irrelevant operators in $D=6$, like $\ff^4$,
near the fixed point in lower dimension: the result is that the
$\ff^4$ term is still irrelevant near four dimension.
Obviously the dimer problem and the Yang-Lee edge singularity problem
are equivalent, i.e. are in the same universality class, until the
dimers have a second order phase transition but for $D<3$ this is no
longer true and they can be different as they actually are: for instance
in $D=1$ we have $\ss_{YL}=-{1\over2}$ and $\aa_{dimer}-1=
"\ss_{dimer}"=1$.

\noindent
There is also another possible approach in order to justify this
result: we could imagine of
replacing $\S$ by $N\S$ in (\ref{z}) and then let $N\rightarrow\infty$,
this is the point of view advocated in (\cite{DG}), for instance.
Performing backward the steps that lead from (\ref{z1}) to (\ref{z}),
we discover
that this substitution is equivalent to perform the
substitutions\footnote{
This is easily understandable looking to the number of derivatives
implied by $-2N\log(z)$ in (\ref{z}) and then looking for the number
of grassman variables needed to have such a number of derivatives
form a formula similar to (\ref{delta}).
}
$\prod_{l\in\G}\,d\eb_l d\ee_l\rightarrow
\prod_{l\in\G}\prod_{A=1}^{A=2N-1}\,d\eb_l^A d\ee_l^A$
and
$\eb\ee\rightarrow\sum_{A=1}^{A=2N-1}\eb^A\ee^A$
in (\ref{z1}) ,where $A$ labels different replica of the original
grassman variables $\eb$ and $\ee$.
This in turn would imply a different HTE expansion: we would get a gas
of branched polymers with loops made of $N$ different types of monomers,
each type being selfavoiding and having a negative activity.
Consequently we would describe very different objects when $N>>1$ and
hence this point of view is not completely satisfactory in this case.

In conclusion we show that there is no contradiction between
$\aa={1\over2}$ and $D_{cr}=6$ and that the apparent paradox is due to a
the fact that the dependence on $\sqrt{\ddt}$ does not cancel in the
free energy differently from what happens in the usual Landau-Ginzburg.



\begin{thebibliography}{99}
\bibitem{Sa}
S. Samuel, \JMaP{21} (1980) 2820
\bibitem{FiRR}
M.E. Fisher, \JMaP{7} (1966) 1776\\
M Rasetti and T. Regge, \RNC{4} (1981) 1
\bibitem{Sh}
Y. Shapir, \JP{A15} (1982) L433
\bibitem{KW}
J.E. Kirkham and D.J. Wallace, \JP{A12} (1979) L47
\bibitem{DG}
E. Br\'ezin, J.C. Guillou and J. Zinn-Justin, Phase transitions and
critical phenomena, editors C. Domb and M.S. Green vol.6
\bibitem{KF}
G.A. Baker and P. Moussa, \JAP{49} (1978) 1360\\
D. Kurze and M.E. Fisher, \PR{B20} (1979) 2785\\
\end{thebibliography}
\end{document}